\documentclass[letterpaper, 10 pt]{ieeeconf}

\usepackage{amsfonts,amsmath,amssymb} 

\DeclareMathOperator{\diag}{diag}         
\def\rbb{\mathbb{R}}
\def\trp{^T}
\def\diag{{\rm diag}}

\def\half{\frac{1}{2}}
\usepackage{psfrag,color}
\usepackage{enumerate,cite,latexsym,graphicx}
\newtheorem{theorem}{Theorem}
\newtheorem{lemma}{Lemma}

\title{\LARGE \bf A Direct Coupling Coherent Quantum Observer for an Oscillatory Quantum Plant}

\author{Ian R.~Petersen 
\thanks{This work was supported by the Air Force Office of Scientific
Research (AFOSR), under agreement number FA2386-16-1-4065. Some of the research presented in this paper was also supported by the Australian Research Council under grant FL110100020.}%
\thanks{Ian R. Petersen is with the Research School of  Engineering, 
        The Australian National University, Canberra ACT 2601, Australia.
         {\tt\small i.r.petersen@gmail.com} } 
}%

\begin{document}

\maketitle
\thispagestyle{empty}
\pagestyle{empty}

\begin{abstract}
A direct coupling coherent observer is constructed for a linear quantum plant which has oscillatory solutions. It is shown that a finite time moving average of the quantum observer output can provide an estimate of the quantum plant output without disturbing this plant signal. By choosing a sufficiently small averaging time and a sufficiently large observer gain, the observer tracking error can be made arbitrarily small. 
\end{abstract}

\section{Introduction} \label{sec:intro}

In recent years, a number of papers have considered the problem of constructing a coherent quantum observer for a linear quantum system \cite{MJP1,VP9a,VP11a,PET14Aa,PET14Ca,PET14Da,PeHun1a,PeHun2a,PeHun3a}. In this problem, the quantum plant is a linear quantum system and the quantum observer is another linear quantum system which is coupled to the quantum plant in some way. Then, the quantum observer is constructed in such a way that it provides an estimate for some of the variables in the quantum plant. 

In the above papers, the quantum plant under consideration is a linear quantum system. In recent years, there has been considerable interest in the modeling and feedback control of linear quantum systems; e.g., see \cite{JNP1,NJP1,ShP5,PET10B}.
Such linear quantum systems commonly arise in the area of quantum optics; e.g., see
\cite{GZ00,BR04}. For such linear quantum system models an important class of quantum control problems are referred to as coherent
quantum feedback control problems; e.g., see \cite{JNP1,NJP1,MaP3,MAB08,ZJ11,VP4,VP5a,HM12,YAM14A,XPD4,XPD2,VuP5}. In these coherent quantum feedback control problems, both the plant and the controller are quantum systems and the controller is typically to be designed to optimize some performance index. The advantage of coherent quantum controllers is that they do not require quantum measurements which inherently lead to the loss of quantum information. The coherent quantum observer problem can be regarded as a special case of the coherent
quantum feedback control problem in which the objective of the observer is to estimate the system variables of the quantum plant. 

In some of the previous papers on quantum observers such as  \cite{MJP1,VP9a,MEPUJ1a}, the coupling between the plant and the observer is via a field coupling.  This enables a one way connection between the quantum plant and the quantum observer. Also, since both the quantum plant and the quantum observer are  open quantum systems, they are both subject to quantum noise. However in the paper \cite{ZJ11}, a coherent quantum control problem is considered in which both field coupling and direct coupling is considered between the quantum plant and the quantum controller. Also,  the papers \cite{PET14Aa,PET14Ca,PET14Da,PeHun1a,PeHun3a} consider  the construction of a coherent quantum observer in which there is only direct coupling between quantum plant and the quantum observer. Furthermore in these papers, both the quantum plant and the quantum observer are assumed to be closed quantum systems which means that they are not subject to quantum noise and are purely deterministic systems.  It is shown  in these papers  that the quantum observer can be constructed to estimate some but not all of the system variables of the quantum plant. However, because of the fact that linear closed quantum systems cannot be asymptotically stable, the observer variables in these papers converge to the plant variables in a time averaged sense.

One significant restriction imposed in the papers \cite{PET14Aa,PET14Ca,PET14Da,PeHun1a,PeHun3a} is that the plant dynamics are such that the plant variables remain constant. In this paper, we investigate whether this restriction can be relaxed and allow for quantum linear plants which have oscillatory solutions. Indeed, the main result of this paper is an extension of the result of \cite{PET14Aa} to the case of a two mode linear quantum plant which is constructed in such a way that oscillatory solutions exist and these can be estimated by a directly coupled quantum observer without disturbing the plant variables of interest. In order to achieve this, we replace the long term time average of the observer output considered in \cite{PET14Aa} with a moving average such that the averaging time is sufficiently short. Then the direct coupled quantum observer is constructed as a linear quantum system using ideas from standard observer theory; e.g., see \cite{Hespanha09}. In this case, the averaged output of the quantum observer does not asymptotically track the plant output but rather it is shown that with a suitably short averaging time and a suitably large observer gain, the observer tracking error can be made arbitrarily small. This result is illustrated with a numerical example. 

\section{Quantum  Systems}
In the  quantum observer network problem under consideration, both the quantum plant and the quantum observer network are linear quantum systems; see also \cite{JNP1,GJ09,ZJ11}. We will restrict attention to closed linear quantum systems which do not interact with an external environment. 
The quantum mechanical behavior of a linear quantum system is described in terms of the system \emph{observables} which are self-adjoint operators on an underlying infinite dimensional complex Hilbert space $\mathfrak{H}$.   The commutator of two scalar operators $x$ and $y$ on ${\mathfrak{H}}$ is  defined as $[x, y] = xy - yx$.~Also, for a  vector of operators $x$ on ${\mathfrak H}$, the commutator of ${x}$ and a scalar operator $y$ on ${\mathfrak{H}}$ is the  vector of operators $[{x},y] = {x} y - y {x}$, and the commutator of ${x}$ and its adjoint ${x}^\dagger$ is the  matrix of operators 
\[ [{x},{x}^\dagger] \triangleq {x} {x}^\dagger - ({x}^\# {x}^T)^T, \]
where ${x}^\# \triangleq (x_1^\ast\; x_2^\ast \;\cdots\; x_n^\ast )^T$ and $^\ast$ denotes the operator adjoint. 

The dynamics of the closed linear quantum systems under consideration are described by non-commutative differential equations of the form
\begin{eqnarray}
\dot x(t) &=& Ax(t); \quad x(0)=x_0
 \label{quantum_system}
\end{eqnarray}
where $A$ is a real matrix in $\rbb^{n
\times n}$, and $ x(t) = [\begin{array}{ccc} x_1(t) & \ldots &
x_n(t)
\end{array}]\trp$ is a vector of system observables; e.g., see \cite{JNP1}. Here $n$ is assumed to be an even number and $\frac{n}{2}$ is the number of modes in the quantum system. 

The initial system variables $x(0)=x_0$ 
are assumed to satisfy the {\em commutation relations}
\begin{equation}
[x_j(0), x_k(0) ] = 2 i \Theta_{jk}, \ \ j,k = 1, \ldots, n,
\label{x-ccr}
\end{equation}
where $\Theta$ is a real skew-symmetric matrix with components
$\Theta_{jk}$.  In the case of a
single quantum harmonic oscillator, we will choose $x=(x_1, x_2)^T$ where
$x_1=q$ is the position operator, and $x_2=p$ is the momentum
operator.  The
commutation relations are  $[q,p]=2 i$.
In general, the matrix $\Theta$ is assumed to be  of the  form
\begin{equation}
\label{Theta}
\Theta=\diag(J,J,\ldots,J)
\end{equation}
 where $J$ denotes the real skew-symmetric $2\times 2$ matrix
$$
J= \left[ \begin{array}{cc} 0 & 1 \\ -1 & 0
\end{array} \right].$$

The system dynamics (\ref{quantum_system}) are determined by the system Hamiltonian  
which is a self-adjoint operator on the underlying  Hilbert space  $\mathfrak{H}$. For the linear quantum systems under consideration, the system Hamiltonian will be a
quadratic form
$\mathcal{H}=\half x(0)\trp R x(0)$, where $R$ is a real symmetric matrix. Then, the corresponding matrix $A$ in 
(\ref{quantum_system}) is given by 
\begin{equation}
A=2\Theta R \label{eq_coef_cond_A}
\end{equation}
 where $\Theta$ is defined as in (\ref{Theta});
e.g., see \cite{JNP1}.
In this case, the  system variables $x(t)$ 
will satisfy the {\em commutation relations} at all times:
\begin{equation}
\label{CCR}
[x(t),x(t)^T]=2{\pmb i}\Theta \ \mbox{for all } t\geq 0.
\end{equation}
That is, the system will be \emph{physically realizable}; e.g., see \cite{JNP1}.

\noindent
{\bf Quantum Plant}

In our proposed direct coupling coherent quantum observer network, the quantum plant is a two mode linear quantum system of the form (\ref{quantum_system}) described by the non-commutative differential equations
\begin{eqnarray}
\dot x_p(t) 
&=& A_px_p(t); \quad x_p(0)=x_{0p}; \nonumber \\
z_p(t) &=& C_px_p(t)
 \label{plant}
\end{eqnarray}
where $z_p(t)$ denotes the vector of system variables to be estimated by the observer network and  $ A_p \in 
\rbb^{4 \times 4}$, $C_p\in \rbb^{1 \times 4}$. 
It is also assumed that this quantum plant corresponds to a plant Hamiltonian
$\mathcal{H}_p=\half x_p(0)\trp R_p x_p(0)$ such that $R_p$ is of the form
\begin{eqnarray}
\label{Rp}
R_p &=& \left[\begin{array}{ll}0 & R_{pc} \\ R_{pc}^T & 0\end{array}\right]; \nonumber \\ 
R_{pc} &=&  \left[\begin{array}{cc} 0 & -\omega_p/2 \\\omega_p/2 & 0 \end{array}\right] = -\frac{\omega_p}{2}J
\end{eqnarray}
where $\omega_p > 0$.
It follows from (\ref{eq_coef_cond_A}) that $A_p=2\Theta_p R_p$ where the matrix $\Theta_p$ is of the form (\ref{Theta}). Hence,
\begin{eqnarray}
\label{Ap}
A_p &=& 2 \left[\begin{array}{ll}J & 0\\0 & J\end{array}\right]\left[\begin{array}{ll}0 & R_{pc} \\ R_{pc}^T & 0\end{array}\right] \nonumber \\
&=& 2\left[\begin{array}{ll}0 & JR_{pc} \\ JR_{pc}^T & 0\end{array}\right] \nonumber \\
&=&  \left[\begin{array}{ll}0 & \omega_p I\\-\omega_p I & 0\end{array}\right].
\end{eqnarray}
From this it follows that the plant equations (\ref{plant}) will have an oscillatory solution. Indeed (\ref{plant}) implies
\[
x_p(t) = e^{A_p t} x_p(0), t \geq 0
\]
where
\[
e^{A_p t} = \left[\begin{array}{ll}I \cos \omega_p t  & I\sin \omega_p t \\ -I\sin \omega_p t  & I \cos \omega_p t \end{array}\right].
\]
Letting $x_p =  \left[\begin{array}{l}x_{p1} \\ x_{p2}\end{array}\right]$, it follows that we can write
\[
\left[\begin{array}{l}x_{p1}(t) \\ x_{p2}(t) \end{array}\right] 
= \left[\begin{array}{l}x_{p1}(0) \cos \omega_p t + x_{p2}(0)\sin \omega_p t \\ 
 -x_{p1}(0)\sin \omega_p t +x_{p2}(0) \cos \omega_p t\end{array}\right].
\]

In addition, we assume that $C_p$ is of the form
\begin{equation}
\label{Cp}
C_p = \left[\begin{array}{ll}C_{p1} & 0 \end{array}\right]
\end{equation}
where $C_{p1}\in \rbb^{1 \times 2}$. Therefore
\[
z_p(t) = C_{p1}x_{p1}(0) \cos \omega_p t + C_{p1}x_{p2}(0)\sin \omega_p t
\]
is also a sinusoidally varying quantity. Furthermore, it follows from (\ref{Ap}) that we can write
\begin{equation}
\label{dzp}
\dot z_p(t) =  \omega_p C_{p1}x_{p2}(t) = \omega_p \tilde z_p(t)
\end{equation}
where $\tilde z_p(t) = C_{p1}x_{p2}(t)$. Furthermore again using (\ref{Ap})
\begin{equation}
\label{dzpt}
\dot{\tilde{z}}_p(t) = C_{p1}\dot x_{p2}(t) = - \omega_p C_{p1}x_{p1}(t) = - \omega_pz_p(t).
\end{equation}

Equations (\ref{dzp}) and (\ref{dzpt}) are the defining equations for $z_p(t)$ and can be written in matrix form
\begin{eqnarray}
\label{zp}
\dot{\bar{z}}_p(t) &=&  \bar A_p \bar z_p(t); \nonumber \\
z_p(t) &=& \bar C_p \bar z_p(t)
\end{eqnarray}
where $\bar z_p = \left[\begin{array}{l}z_p \\ \tilde z_p \end{array}\right]$, $\bar A_p = \omega_p J$ and 
$\bar C_p = \left[\begin{array}{ll}1 & 0 \end{array}\right]$. 

The sinusoidal form of the quantity to be estimated $z_p(t)$ will apply if the plant is not coupled to the observer. However, when the plant is coupled to the quantum observer, this may no longer be the case. We will show that if the quantum observer is suitably designed, the plant quantity to be estimated  $z_p(t)$ will be unaffected by the presence of the observer.

\noindent
{\bf Quantum Observer}

We now describe a single quantum harmonic oscillator system  which will correspond to the quantum observer; see also \cite{JNP1,GJ09,ZJ11}. 
This system is described by a non-commutative differential equation of the form
\begin{eqnarray}
\dot x_o(t) &=& A_ox_o(t);\nonumber \\
z_o(t) &=& C_ox_o(t)
 \label{observer}
\end{eqnarray}
where the observer output $z_o$ is the observer estimate variable and $C_o\in \rbb^{1\times 2}$. Also, $A_o \in \rbb^{2
\times 2}$.
We assume that the plant variables commute with the observer variables. The system dynamics (\ref{observer}) are determined by the system Hamiltonian 
which is a self-adjoint operator on the underlying infinite dimensional Hilbert space for the system $\mathfrak{H}_o$. For the single quantum harmonic oscillator system under consideration, the system Hamiltonian is given by the
quadratic form
$\mathcal{H}_o=\half x(0)\trp R_o x(0)$, where $R_o$ is a real symmetric matrix. Then, the corresponding matrix $A_o$ in 
(\ref{observer}) is given by 
\begin{equation}
A_o=2 J R_o \label{eq_coef_cond_Ao}.
\end{equation}

In our proposed direct coupling coherent quantum observer, the quantum plant (\ref{plant}) will be directly coupled to the coherent quantum observer (\ref{observer}) by introducing a coupling Hamiltonian
\begin{equation}
\label{coupling_hamiltonian}
\mathcal{H}_c=\half x_p(0)\trp R_c x_o(0)+\half x_o(0)\trp R_c\trp x_p(0)
\end{equation}
where $R_c\in \rbb^{4\times 2}$.

The augmented quantum linear system consisting of the quantum plant and the direct coupled  quantum observer is then a quantum system of the form (\ref{quantum_system}) described by the total Hamiltonian
\begin{eqnarray}
\mathcal{H}_a &=& \mathcal{H}_p+\mathcal{H}_c+\mathcal{H}_o\nonumber \\
 &=& \half x_a\trp R_a x_a
\label{total_hamiltonian}
\end{eqnarray}
where
$x_a = \left[\begin{array}{l}x_p\\x_o\end{array}\right]$ and 
$R_a = \left[\begin{array}{ll}R_p & R_c\\R_c^T & R_o\end{array}\right]$. Then, using (\ref{eq_coef_cond_A}), it follows that the augmented quantum linear system is described by the equations
\begin{eqnarray}
\left[\begin{array}{l}\dot x_p(t)\\\dot x_o(t)\end{array}\right] &=& 
A_a\left[\begin{array}{l} x_p(t)\\ x_o(t)\end{array}\right];~ x_p(0)=x_{0p};~ x_o(0)=x_{0o};\nonumber \\
z_p(t) &=& C_px_p(t);\nonumber \\
z_o(t) &=& C_ox_o(t)
\label{augmented_system}
\end{eqnarray}
where $A_a = 2\Theta_a R_a$. Here 
\[
\Theta_a =\left[\begin{array}{ll}\Theta_p & 0\\0 & J \end{array}\right].
\]

We wish to construct the quantum observer so that the time averaged quantity $\frac{1}{T}\int_{t-T}^{t}z_o(\tau)d\tau$ provides a good approximation to the quantity $z_p(t)$ for a suitable choice of the averaging time $T$.

\section{Constructing  the Direct Coupling Coherent Quantum Observer}
We now describe the construction of the  direct coupled linear quantum observer.  
We suppose that the matrices $R_o$,  $R_c$, $C_o$ are such that 
\begin{equation}
\label{design1}
 R_c =  \alpha\beta^T,~~ \alpha =   C_{p}^T,~~ R_o > 0
\end{equation}
where  $\beta \in \rbb^{2\times 1}$. Therefore, it follows from (\ref{Cp}) that
\[
  R_a = \left[\begin{array}{lll}0 & -\omega_p J/2 & \tilde \alpha \beta^T \\
                \omega_p J/2 & 0 & 0\\
                \beta \tilde \alpha^T & 0 & R_o \end{array}\right]
\]
where $\tilde \alpha = C_{p1}^T$. 
Hence, the augmented system equations (\ref{augmented_system}) describing the combined plant-observer system become
\begin{eqnarray}
  \dot x_{p1}(t)&=& \omega_p x_{p2}(t)  + 2J \tilde \alpha\beta^Tx_o(t);\nonumber \\
  \dot x_{p2}(t)&=& -\omega_p x_{p1}(t);\nonumber \\
\dot x_o(t)&=&2 J \beta z_{p}(t)+2  J R_ox_o(t);\nonumber \\
z_p(t) &=&  C_{p1} x_{p1}(t);\nonumber \\
z_o(t) &=& C_ox_o(t). 
\label{augmented_system1}
\end{eqnarray}
It follows that
\begin{eqnarray}
\label{dzpa}
  \dot z_p(t) &=&  \omega_p C_{p1} x_{p2}(t)  + 2\tilde \alpha^T J \tilde \alpha\beta^Tx_o(t) \nonumber \\
              &=& \omega_p C_{p1} x_{p2}(t) \nonumber \\
& = &  \omega_p \tilde z_p(t)
\end{eqnarray}
since $\tilde \alpha^T J \tilde \alpha = 0$. Furthermore, 
\begin{equation}
\label{dzpta}
\dot{\tilde{z}}_p(t) =   C_{p1} \dot x_{p2}(t) = -\omega_p C_{p1}  x_{p1}(t) =  -\omega_p z_p(t).
\end{equation}
Equations (\ref{dzpa}) and (\ref{dzpta}) are the same as equations  (\ref{dzp}) and (\ref{dzpt}). That is, when the quantum observer is connected to the quantum plant, the  equations describing $z_p(t)$ are not changed. 

Now in order to construct suitable values for the quantities $\beta$ and $R_o$, we note that we can write down equations for augmented system involving only the variables $\bar z_p(t)$ and $x_o(t)$ as follows:
\begin{eqnarray}
\label{augmented_system2}
\dot{\bar{z}}_p(t) &=&  \bar A_p \bar z_p(t); \nonumber \\
\dot x_o(t)&=&2  J R_ox_o(t)+2 J \beta \bar C_p \bar z_{p}(t);\nonumber \\
z_o(t) &=& C_ox_o(t); \nonumber \\
z_p(t) &=& \bar C_p \bar z_p(t). 
\end{eqnarray}
These equations are of the form 
\begin{eqnarray}
\label{augmented_system3}
\left[\begin{array}{l}\dot{\bar{z}}_p(t)\\\dot x_o(t)\end{array}\right] 
&=& \bar A_a \left[\begin{array}{l}\bar{z}_p(t)\\ x_o(t)\end{array}\right];  \nonumber \\
z_o(t)&=& \left[\begin{array}{cc} 0 & C_o\end{array}\right]\left[\begin{array}{l}\bar{z}_p(t)\\ x_o(t)\end{array}\right];  \nonumber \\
z_p(t) &=& \left[\begin{array}{cc} \bar C_p & 0 \end{array}\right] \left[\begin{array}{l}\bar{z}_p(t)\\ x_o(t)\end{array}\right]
\end{eqnarray}
where
\[
 \bar A_a = \left[\begin{array}{cc}\bar A_p & 0 \\
2 J \beta \bar C_p &  A_o \end{array}\right].
\]
Furthermore, the equations (\ref{augmented_system2}) are in the form of the standard plant-observer equations if we choose $R_o > 0$ such that 
\begin{equation}
\label{Ro}
A_o = 2  J R_o = \bar A_p - L \bar C_p; \quad L = 2 J \beta
\end{equation}
where $L \in \rbb^{2\times 1}$ is the observer gain; e.g., see \cite{Hespanha09}. Then, letting $e(t) = x_o(t) - \bar z_p(t)$, it follows that
\begin{eqnarray}
\label{de}
\dot e(t) &=& \dot x_o(t) - \dot{\bar{z}}_p(t) \nonumber \\
&=& \bar A_p x_o(t) - L \bar C_p x_o(t) + L C_p \bar z_{p}(t) 
-  \bar A_p \bar z_p(t) \nonumber \\
&=& \left(\bar A_p - L \bar C_p\right)e(t) \nonumber \\
&=& A_o e(t).
\end{eqnarray}
Now it follows from (\ref{Ro}) that
\[
2  J R_o = \omega_p J - 2 J \beta \bar C_p
\]
and hence,
\[
R_o = \frac{\omega_p}{2}  I - \beta \bar C_p.
\]
However, we require that $R_o$ is symmetric and positive-definite. Hence, we choose $\beta$ to be of the form
\begin{equation}
\label{beta}
\beta =  -\mu \bar C_p^T = \left[\begin{array}{l}-\mu\\0\end{array}\right]
\end{equation}
where $\mu > 0$. Therefore
\begin{equation}
\label{Ro1}
R_o = \frac{\omega_p}{2}  I + \left[\begin{array}{ll}\mu & 0\\0 & 0\end{array}\right] 
=  \left[\begin{array}{ll}\mu+ \frac{\omega_p}{2}  & 0\\0 &  \frac{\omega_p}{2}\end{array}\right] > 0.
\end{equation}
This defines $A_o$ in (\ref{observer}) as
\[
A_o = 2JR_o = \left[\begin{array}{cc}0  & \omega_p\\ -2\mu- \omega_p&  0\end{array}\right].
\]
Also, since we want $z_o(t)$ to provide an estimate of $z_p(t)$, we choose
\begin{equation}
\label{Co}
C_o = \bar C_p = \left[\begin{array}{ll}1 & 0 \end{array}\right].
\end{equation}

We now calculate the averaged value of the estimation error 
$z_o(t)-z_p(t) =  C_o e(t)$. It follows from  (\ref{de}) that we can write the averaged value of the estimation error in the form 
\begin{equation}
\label{averr0}
\frac{1}{T} \int_{t-T}^t\left(z_o(\tau)-z_p(\tau)\right)d \tau =  g_1(t) e_1(0)+g_2(t) e_2(0)
\end{equation}
for all $t \geq T$.
Indeed, it follows from (\ref{de})
and the fact that $A_o$ is nonsingular that 
\begin{eqnarray}
\label{averr1}
\lefteqn{\frac{1}{T} \int_{t-T}^t\left(z_o(\tau)-z_p(\tau)\right)d \tau}\nonumber \\
&=& \frac{1}{T} \int_{t-T}^t C_oe(\tau) d \tau  \nonumber \\
 &=& \frac{1}{T} \int_{t-T}^t C_oe^{A_o \tau}e(0) d \tau  \nonumber \\
&=& \frac{C_o}{T}e^{A_o t}\left(I-e^{-A_o T}\right) A_o^{-1}e(0)
\end{eqnarray}
for all $t \geq T$. Also, we can calculate
\begin{eqnarray*}
e^{A_o t} &=& \left[\begin{array}{cc} \cos \omega_o t & \frac{\omega_p}{\omega_o} \sin \omega_o t \\
-\frac{2\mu+ \omega_p}{\omega_o} \sin \omega_o t & \cos \omega_o t \end{array}\right]; \nonumber \\
A_o^{-1} &=& \left[\begin{array}{cc} 0 & \frac{-1}{2\mu+ \omega_p} \\
\frac{1}{\omega_p} & 0 \end{array}\right]
\end{eqnarray*}
where $\omega_o = \sqrt{\omega_p\left(2\mu+ \omega_p\right)} > 0$. Hence using (\ref{averr1}), we calculate
\begin{eqnarray}
\label{averr2}
\lefteqn{\frac{1}{T} \int_{t-T}^t\left(z_o(\tau)-z_p(\tau)\right)d \tau}\nonumber \\
&=& \frac{C_o}{T}\left[\begin{array}{cc} \cos \omega_o t & \frac{\omega_p}{\omega_o} \sin \omega_o t \\
-\frac{2\mu+ \omega_p}{\omega_o} \sin \omega_o t & \cos \omega_o t \end{array}\right]\nonumber \\
&& \times \left[\begin{array}{cc} 1-\cos \omega_o T & \frac{\omega_p}{\omega_o} \sin \omega_o T \\
-\frac{2\mu+ \omega_p}{\omega_o} \sin \omega_o T & 1-\cos \omega_o T \end{array}\right] \nonumber \\
&& \times \left[\begin{array}{cc} 0 & \frac{-1}{2\mu+ \omega_p} \\
\frac{1}{\omega_p} & 0 \end{array}\right]e(0) \nonumber \\
&=& \frac{1}{T}\left[\begin{array}{cc} \cos \omega_o t & \frac{\omega_p}{\omega_o} \sin \omega_o t 
\end{array}\right]\nonumber \\
 && \times \left[\begin{array}{cc} \frac{1}{\omega_o} \sin \omega_o T & -\frac{1-\cos \omega_o T}{2\mu+ \omega_p} \\
  \frac{1-\cos \omega_o T}{\omega_p} & \frac{1}{\omega_o} \sin \omega_o T \end{array}\right] e(0) \nonumber \\
&=& 
\left(\cos \omega_o t\sin \omega_o T+ \sin \omega_o t \left(1-\cos \omega_o T\right)\right) \frac{e_1(0)}{T\omega_o} \nonumber \\
&&+\left(\frac{\omega_p}{\omega_o^2}\sin \omega_o t \sin \omega_o T
- \frac{\cos \omega_o t \left(1-\cos \omega_o T\right)}{2\mu+ \omega_p}\right)\nonumber \\
&&\times \frac{e_2(0)}{T} \nonumber \\
&=& g_1(t) e_1(0)+g_2(t) e_2(0)
\end{eqnarray}
for all $t \geq T$. 

Also, we note that $\omega_o \rightarrow \infty$ as $\mu \rightarrow \infty$. Then, using the formula (\ref{averr2}), we obtain the following lemma. 

\begin{lemma}
\label{L1}
Consider the quantum plant and quantum observer constructed above. Then for any $\epsilon >0$  and any averaging time $T > 0$, there exits a constant $\mu  >0$ defining the observer gain such that the average estimation error given in (\ref{averr0}) satisfies
\[
g_1(t)^2+g_2(t)^2 \leq \epsilon
\]
for all $t \geq T$. 
\end{lemma}

This lemma shows that given any averaging time $T > 0$, we can always find an observer with an arbitrarily small averaged estimation error (\ref{averr0}). 

We now show that for sufficiently small $T > 0$, the quantity $\frac{1}{T} \int_{t-T}^tz_p(\tau)d \tau$ will provide a good approximation to $z_p(t)$. It follows from  (\ref{zp}), (\ref{augmented_system2}) that we can write the difference between the averaged plant output and the plant output in the form 
\begin{equation}
\label{avzp}
\frac{1}{T} \int_{t-T}^tz_p(\tau)d \tau - z_p(t) =  h_1(t) z_p(0)+h_2(t) \tilde z_p(0)
\end{equation}
for all $t \geq T$.
Indeed, it follows from (\ref{zp}), (\ref{augmented_system2}) and the fact that $\bar A_p$ is nonsingular that
\begin{eqnarray}
\label{avzperr}
\lefteqn{\frac{1}{T} \int_{t-T}^tz_p(\tau)d \tau - z_p(t)} \nonumber \\
&=& \frac{1}{T} \int_{t-T}^t\bar C_p e^{\bar A_p \tau}z_p(0)d \tau -\bar C_p e^{\bar A_p t}\bar z_p(0)\nonumber \\
&=& \frac{\bar C_p}{T}e^{\bar A_p t}\left(I-e^{-\bar A_p T}\right) \bar A_p^{-1}\bar z_p(0)
-\bar C_p e^{\bar A_p t}\bar z_p(0)\nonumber \\
&=& \bar C_pe^{\bar A_p t} \left(\frac{\bar A_p^{-1}}{T}- \frac{e^{-\bar A_p T}A_p^{-1}}{T} -I \right)\bar z_p(0)
\end{eqnarray}
for all $t \geq T$. Also, we can calculate
\begin{eqnarray*}
e^{\bar A_p t} &=& \left[\begin{array}{cc} \cos \omega_p t & \sin \omega_p t \\
-\sin \omega_p t & \cos \omega_p t \end{array}\right]; \nonumber \\
A_o^{-1} &=& \left[\begin{array}{cc} 0 & \frac{-1}{\omega_p} \\
\frac{1}{\omega_p} & 0 \end{array}\right].
\end{eqnarray*}
Hence using (\ref{avzp}), we calculate
\begin{eqnarray}
\label{avzperr1}
\lefteqn{\frac{1}{T} \int_{t-T}^tz_p(\tau)d \tau - z_p(t)} \nonumber \\
&=& \bar C_pe^{\bar A_p t} \left(\frac{\bar A_p^{-1}}{T}- \frac{e^{-\bar A_p T}A_p^{-1}}{T} -I \right)\bar z_p(0)\nonumber \\
&=& \left[\begin{array}{cc} \cos \omega_p t & \sin \omega_p t  \end{array}\right] \nonumber \\
&&\times \left(\begin{array}{cc}1 - \frac{1}{\omega_p T} \sin \omega_p T & - \frac{1}{\omega_p T}\left(1-\cos \omega_p T\right) \\
\frac{1}{\omega_p T}\left(1-\cos \omega_p T\right) & 1 - \frac{1}{\omega_p T} \sin \omega_p T
\end{array}\right)\nonumber \\
&& \times \bar z_p(0)\nonumber \\
&=& h_1(t) z_p(0)+h_2(t) \tilde z_p(0)
\end{eqnarray}
for all $t \geq T$. From this formula, we obtain the following Lemma. 
\begin{lemma}
\label{L2}
Consider the quantum plant defined as above. Then for any $\epsilon >0$, there exits an averaging time $T > 0$ such that the difference between the averaged plant output and the plant output given in (\ref{avzp}) satisfies
\[
h_1(t)^2+h_2(t)^2 \leq \epsilon
\]
for all $t \geq T$. 
\end{lemma}

Combining Lemma \ref{L1} and Lemma \ref{L2}, we obtain the following theorem which is the main result of this paper. 
\begin{theorem}
\label{T1}
Consider a quantum plant described by equations (\ref{plant}), (\ref{Rp}), (\ref{Ap}), (\ref{Cp}) and a quantum observer described by equations (\ref{observer}), (\ref{design1}), (\ref{beta}), (\ref{Ro1}), (\ref{Co}). Then for any $\epsilon >0$  there exists an  averaging time $T > 0$ and a constant $\mu  >0$ defining the observer gain such that the  difference between the averaged observer output and the plant output is of the form 
\begin{eqnarray*}
\lefteqn{\frac{1}{T} \int_{t-T}^tz_o(\tau)d \tau - z_p(t)}\nonumber \\
&=& g_1(t) e_1(0)+g_2(t) e_2(0)+ h_1(t) z_p(0)+h_2(t) \tilde z_p(0)\nonumber \\
&=& g_1(t) x_{o1}(0) +  g_2(t)x_{o2 }(0)\nonumber \\
&&+ \left(h_1(t)-g_1(t)\right) z_p(0)+ \left(h_2(t)-g_2(t)\right) \tilde z_p(0)
\end{eqnarray*}
where 
\[
g_1(t)^2+g_2(t)^2 +h_1(t)^2+h_2(t)^2\leq \epsilon
\]
for all $t \geq T$. 
\end{theorem}
This theorem shows that we can always construct a direct coupled quantum observer and corresponding averaging time $T > 0$ such that the averaged output of the direct coupled quantum observer is arbitrarily close to the  output of the plant to be estimated.

\section{Illustrative Example}
We now present an example to illustrate the direct coupled quantum observer described in the previous section. We consider a quantum plant which is a modification of the example considered in \cite{PET14Aa} to allow for 
an oscillatory plant with a nonlinear Hamiltonian. In particular, we consider a quantum plant of the form described by equations (\ref{plant}), (\ref{Rp}), (\ref{Ap}), (\ref{Cp})  with $\omega_p =1$ and $C_{p1} = [1~0]$.

Also, we consider a quantum observer defined by equations (\ref{observer}), (\ref{design1}), (\ref{beta}), (\ref{Ro1}), (\ref{Co}) with $\mu > 0$ to be specified.  Then the corresponding augmented plant-observer system can be described by the equations (\ref{augmented_system3}). 

\begin{figure}[htbp]
\begin{center}
\includegraphics[width=8.5cm]{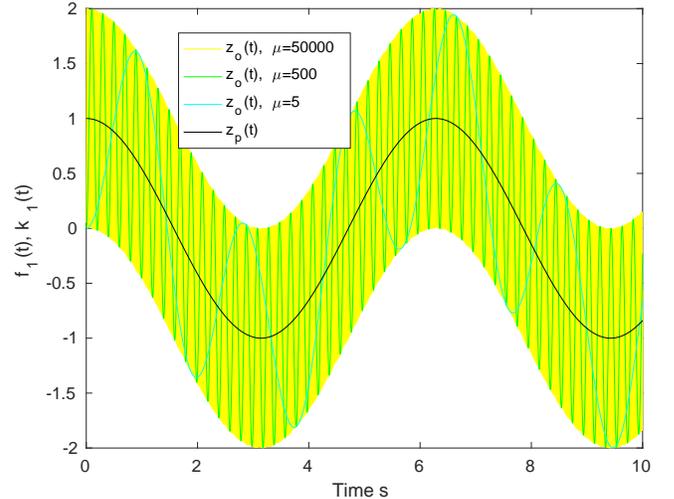}
\end{center}
\caption{Coefficient functions $f_1(t)$ and $k_1(t)$ defining $z_p(t)$ and $z_o(t)$.}
\label{F1a}
\end{figure}

\begin{figure}[htbp]
\begin{center}
\includegraphics[width=8.5cm]{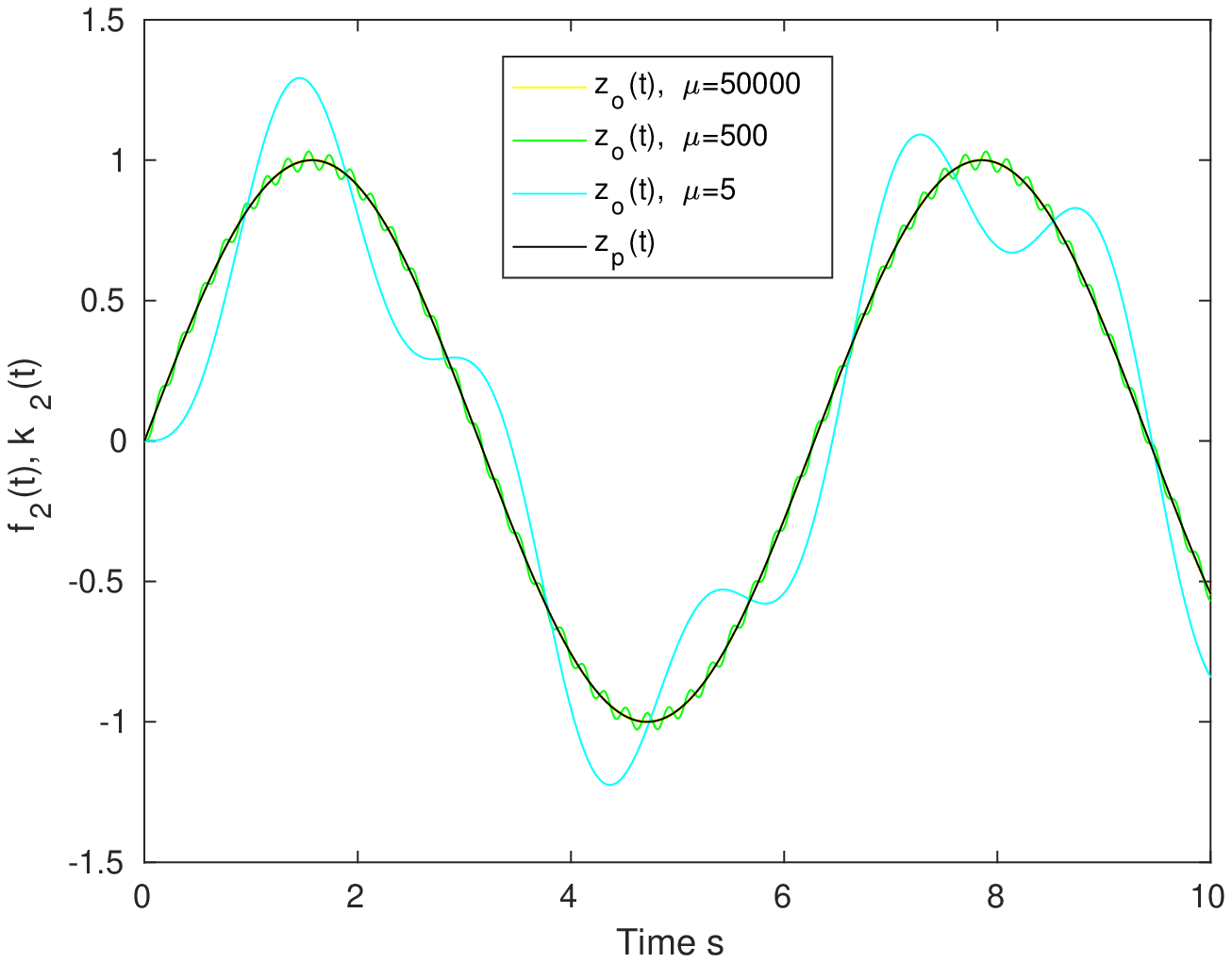}
\end{center}
\caption{Coefficient functions $f_2(t)$ and $k_2(t)$ defining $z_p(t)$ and $z_o(t)$.}
\label{F1b}
\end{figure}

\begin{figure}[htbp]
\begin{center}
\includegraphics[width=8.5cm]{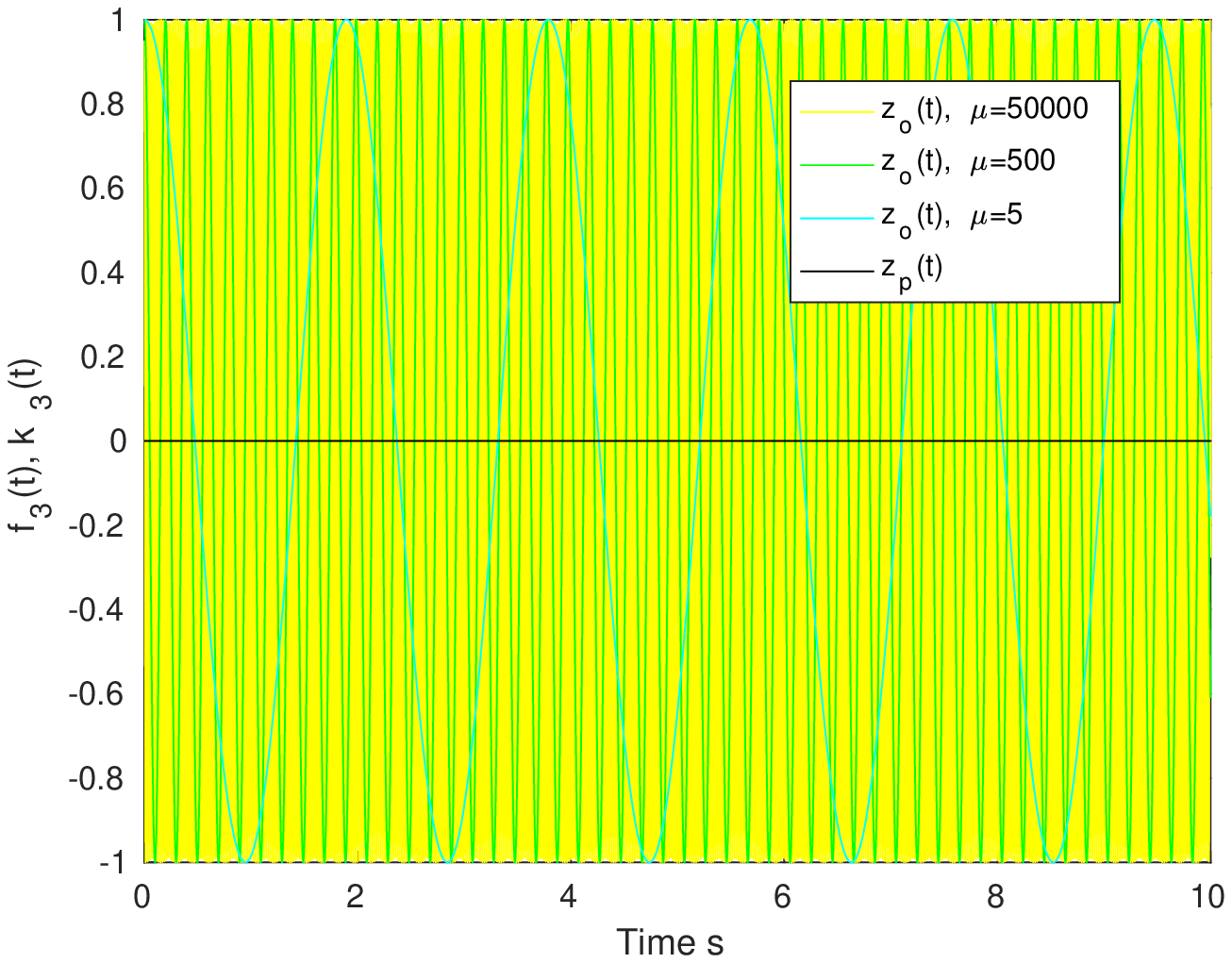}
\end{center}
\caption{Coefficient functions $f_3(t) \equiv 0$ and $k_3(t)$ defining $z_p(t)$ and $z_o(t)$.}
\label{F1c}
\end{figure}

\begin{figure}[htbp]
\begin{center}
\includegraphics[width=8.5cm]{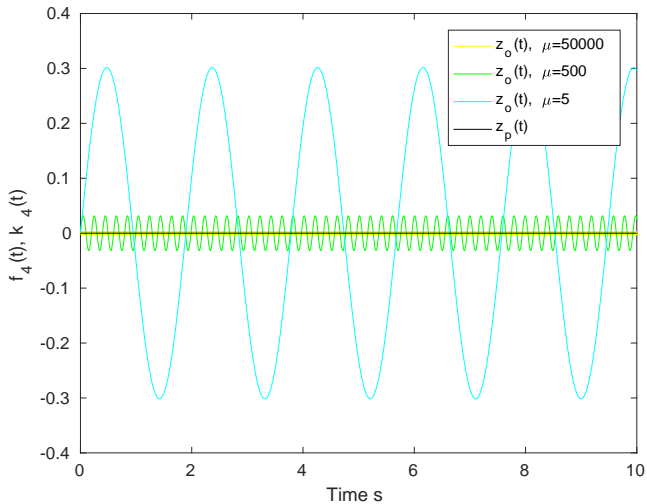}
\end{center}
\caption{Coefficient functions $f_4(t) \equiv 0$ and $k_4(t)$ defining $z_p(t)$ and $z_o(t)$.}
\label{F1d}
\end{figure}

Now it follows from (\ref{augmented_system3}) that the plant output can be written in the form
\[
z_p(t) =  \bar C_p e^{\bar A_p t}\bar z_p(0) = f_1(t) z_p(0) + f_2(t) \tilde z_p(0)
\]
and the observer output can be written in the form
\begin{eqnarray*}
z_o(t) &=& \left[\begin{array}{cc} 0 & C_o\end{array}\right] e^{\bar A_a t}\left[\begin{array}{l}\bar{z}_p(0)\\ x_o(0)\end{array}\right]\nonumber \\
 &=& k_1(t) z_p(0) + k_2(t) \tilde z_p(0) \nonumber \\
&&+  k_3(t) x_{o1}(0) +  k_4(t) x_{o2}(0).
\end{eqnarray*}
In simulating the quantum plant and observer system, we cannot plot the quantities $z_p(t)$, $z_o(t)$ since these are operator functions of time. However, we can plot the real quantities $f_1(t)$,  $f_2(t)$, $k_1(t)$,  $k_2(t)$, $k_3(t)$,  $k_4(t)$. The plots of these quantities are shown in Figures \ref{F1a}-\ref{F1d} for $\mu = 5$, $\mu = 500$, $\mu = 50000$. 

From these figures, we can see that there is no overall improvement in the observer estimation as we vary the constant $\mu$. Hence, as indicated by the theory developed above, we consider the averaged observer output $z_{oav}(t)$ with an averaging time of $T=0.1$. In order to plot this quantity over the entire time interval being considered, we define
\[
z_{oav}(t) = \left\{\begin{array}{ll}\frac{1}{t} \int_{0}^tz_o(\tau)d \tau & \mbox{ for } t \in [0,T); \\
\frac{1}{T}\int_{t-T}^tz_o(\tau)d \tau & \mbox{ for } t \geq T. 
\end{array}\right.
\]
Then as above, this quantity can be written in the form 
\begin{eqnarray*}
z_{oav}(t)&=&\frac{1}{t} \int_{0}^t \left[\begin{array}{cc} 0 & C_o\end{array}\right] e^{\bar A_a \tau}\left[\begin{array}{l}\bar{z}_p(0)\\ x_o(0)\end{array}\right] d \tau  \nonumber \\
&=& \frac{\left[\begin{array}{cc} 0 & C_o\end{array}\right]}{t}\left(e^{\bar A_p t}-I\right) \bar A_p^{-1}
\left[\begin{array}{l}\bar{z}_p(0)\\ x_o(0)\end{array}\right]\nonumber \\
&=& l_1(t) z_p(0) + l_2(t) \tilde z_p(0) \nonumber \\
&&+  l_3(t) x_{o1}(0) +  l_4(t) x_{o2}(0)
\end{eqnarray*}
for $t \in [0,T)$ and 
\begin{eqnarray*}
z_{oav}(t)&=&\frac{1}{T} \int_{t-T}^t \left[\begin{array}{cc} 0 & C_o\end{array}\right] e^{\bar A_a \tau}\left[\begin{array}{l}\bar{z}_p(0)\\ x_o(0)\end{array}\right] d \tau  \nonumber \\
&=& \frac{\left[\begin{array}{cc} 0 & C_o\end{array}\right]}{T}e^{\bar A_p t}\left(I-e^{-\bar A_p T}\right) \bar A_p^{-1}
\left[\begin{array}{l}\bar{z}_p(0)\\ x_o(0)\end{array}\right]\nonumber \\
&=& l_1(t) z_p(0) + l_2(t) \tilde z_p(0) \nonumber \\
&&+  l_3(t) x_{o1}(0) +  l_4(t) x_{o2}(0)
\end{eqnarray*}
for $t \geq T$. 

\begin{figure}[htbp]
\begin{center}
\includegraphics[width=8.5cm]{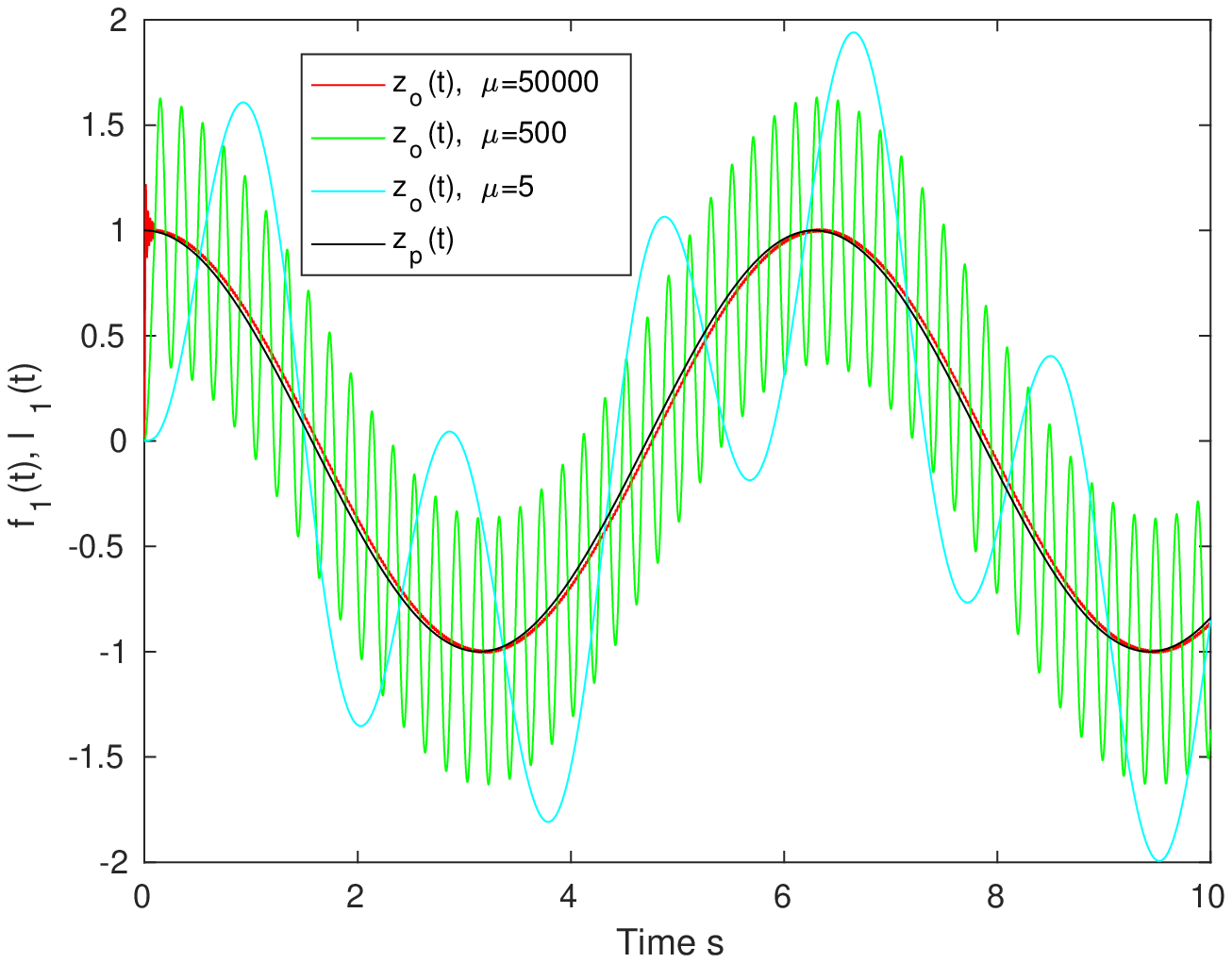}
\end{center}
\caption{Coefficient functions $f_1(t)$ and $l_1(t)$ defining $z_p(t)$ and $z_{oav}(t)$.}
\label{F2a}
\end{figure}

\begin{figure}[htbp]
\begin{center}
\includegraphics[width=8.5cm]{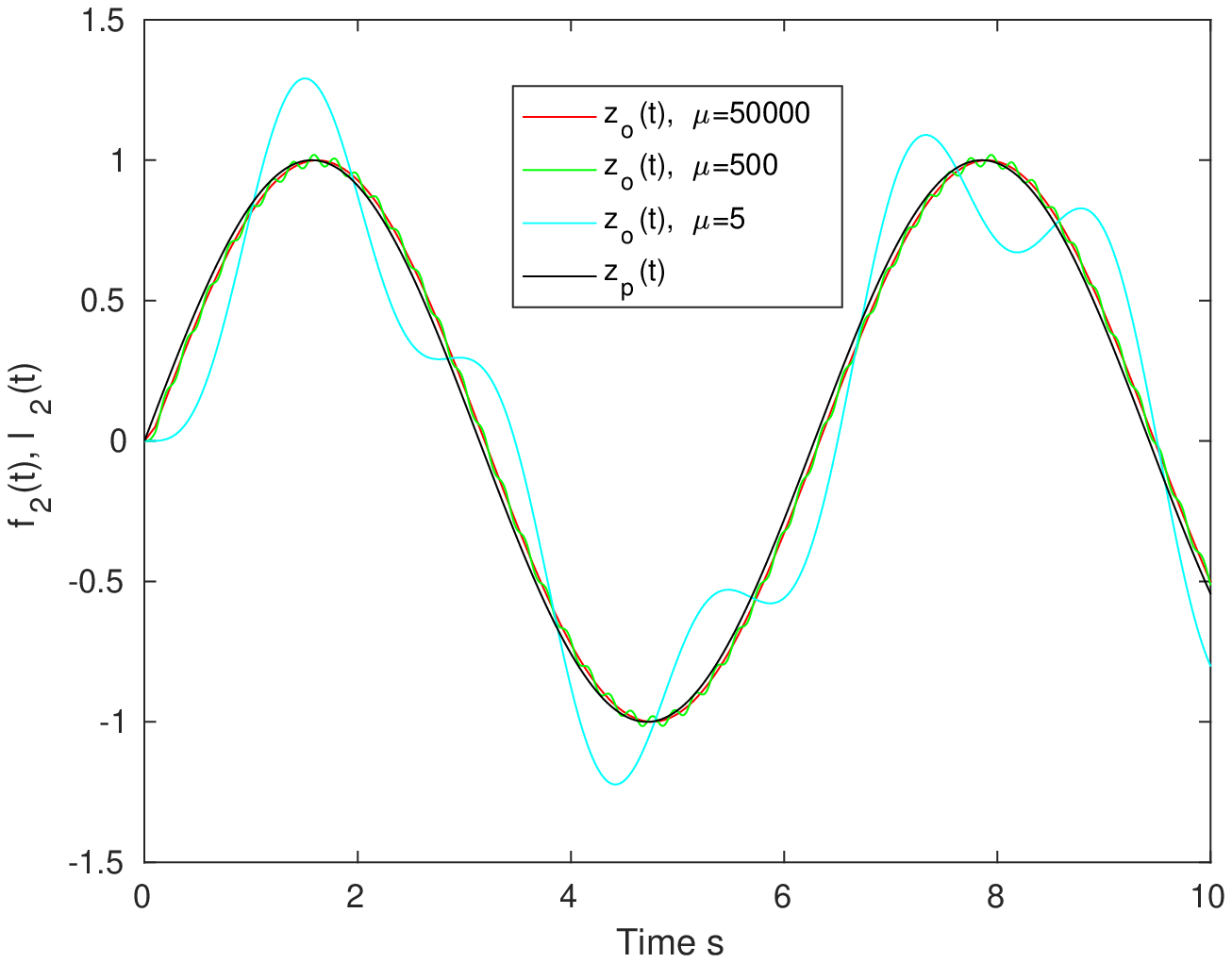}
\end{center}
\caption{Coefficient functions $f_2(t)$ and $l_2(t)$ defining $z_p(t)$ and $z_{oav}(t)$.}
\label{F2b}
\end{figure}

\begin{figure}[htbp]
\begin{center}
\includegraphics[width=8.5cm]{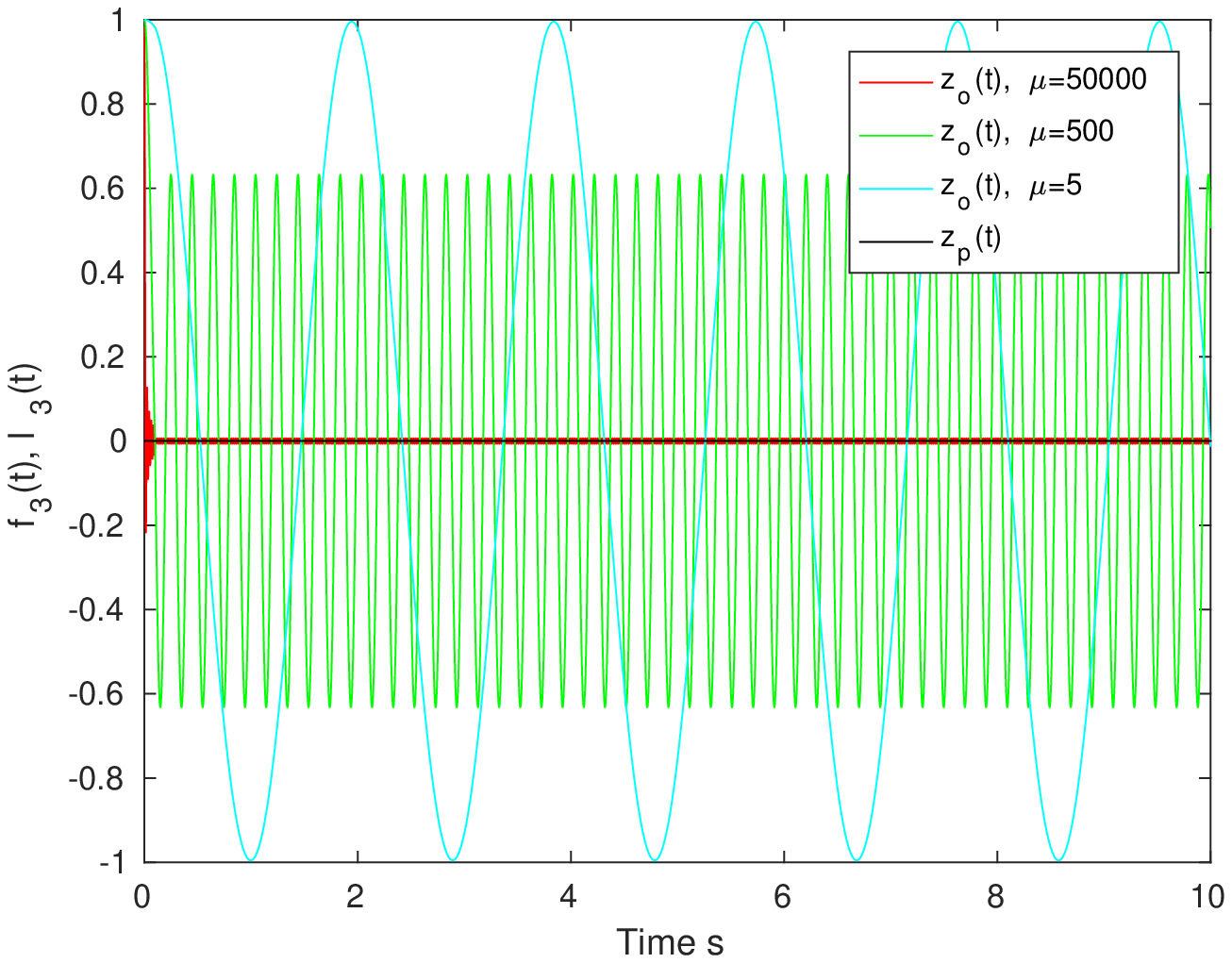}
\end{center}
\caption{Coefficient functions $f_3(t)\equiv 0$ and $l_3(t)$ defining $z_p(t)$ and $z_{oav}(t)$.}
\label{F2c}
\end{figure}

\begin{figure}[htbp]
\begin{center}
\includegraphics[width=8.5cm]{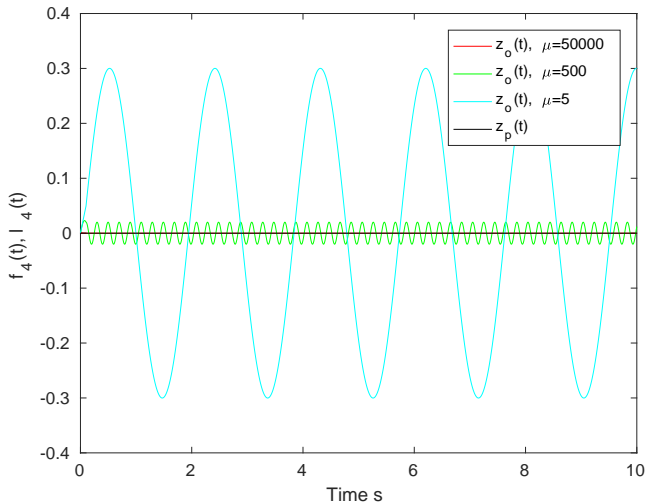}
\end{center}
\caption{Coefficient functions $f_4(t)\equiv 0$ and $l_4(t)$ defining $z_p(t)$ and $z_{oav}(t)$.}
\label{F2d}
\end{figure}

Then we plot the quantities $f_1(t)$,  $f_2(t)$, $l_1(t)$,  $l_2(t)$, $l_3(t)$,  $l_4(t)$ as shown in Figures \ref{F2a}-\ref{F2d} for $\mu = 5$, $\mu = 500$, $\mu = 50000$. 

These figures show that for the given value of $T$, increasing the value of the parameter $\mu$ leads to the averaged observer output signal $z_{oav}(t)$ providing an improved estimate of the plant output signal $z_p(t)$ as expected from the theory derived in the previous section. 

\section{Conclusions}
In this paper, we have described the construction of a direct coupling coherent quantum observer to provide an estimate of an output of a given quantum plant which exhibits oscillatory behaviour. The quantum observer estimate is a finite time average of the observer output signal. The main result of the paper shows that if the averaging time is made sufficiently small and a parameter $\mu$ defining the observer gain is made sufficiently large, then an arbitrarily small estimation error can be achieved. 


\end{document}